\documentstyle[12pt]{article}
\begin{document}
\begin{center}
{\bf Multiparticle production and perturbative QCD}

\bigskip

I.M. Dremin

\bigskip

{\it Lebedev Physical Institute, Moscow 119991, Russia}

\end{center}

\begin{abstract}
The perturbative quantum chromodynamics (QCD) is quite successful in the
description of main features of multiparticle production processes. Ten most 
appealing characteristics are described in this brief review talk and compared
with QCD predictions. The general perturbative QCD approach is demonstrated 
and its problems are discussed. It is shown that the analytical calculations
at the parton level with the low-momentum cut-off reproduce experimental data
on the hadronic final state surprisingly accurately even though the perturbative
expansion parameter is not very small. Moreover, the perturbative QCD has been
able not only to {\it describe} the existing data but also to {\it predict}
many bright qualitative phenomena.
\end{abstract}

\begin{center}
CONTENTS
\end{center}

\noindent 1. Introduction\\
2. QCD equations\\
3. Comparison with experiment\\
3.1. The energy dependence of mean multiplicity\\
3.2. Oscillations of cumulant moments\\
3.3. Difference between quark and gluon jets\\
3.4. The hump-backed plateau\\
3.5. Difference between heavy- and light-quark jets\\
3.6. Color coherence in 3-jet events\\
3.7. Intermittency and fractality\\
3.8. The energy behavior of higher moments of multiplicity distributions\\
3.9. Subjet multiplicities\\
3.10.Jet universality\\
4. Conclusions and outlook\\

\section{Introduction}

Multiparticle production is the main process of very high energy particle
interactions. Studying it, one hopes to get knowledge on validity of our
general ideas about the structure of the matter at smallest distances, on
new states of matter which could be created at these extreme conditions, 
on asymptotical properties of strong interactions, on confinement etc.
One should understand these processes also to separate the signals for new
physics from the conventional QCD background.

Theoretical interpretation of these processes evolved from statistical and
hydrodynamical approaches to multiperipheral models and QCD. Many models
have been elaborated and their computerized Monte Carlo versions are 
available for a detailed comparison with experimental data. In particular,
A. Capella told us about the dual parton model proposed and developed in
Orsay in the late 1970s by a group of theorists including J. Tran Thanh Van.
Neither these phenomenological models nor QCD approach can evade a problem
of transition from partons (quarks, gluons) to the observed particles. This
stage is treated phenomenologically in any of them. It introduces additional
parameters which can give us a hint to the confinement property, but they are
sometimes hard to control.

The whole process is considered as a cascade of consecutive emissions of
partons each of which produces observed hadrons in a jet-like manner.
Jets from primary quarks were discovered in 1975 with the angular distribution
expected for spin 1/2 quarks. Gluons emitted by quarks at large transverse
momenta can be described by perturbative QCD due to the asymptotical freedom
property, and such processes are used to determine the value of the coupling
strength. However, one can try to proceed to lower transverse momenta when
many jets (and consequently many hadrons) are created. These processes are 
of main concern for this survey.

Perturbative QCD analysis can be directly applied to $e^{+}e^{-}$-data where
the initial state is determined by quark jets or to jets separated from
final states in $ep, p\overline{p}$ etc processes.

Usually, Monte Carlo models deal with matrix elements (actually, probabilities)
of a process at the parton level plus hadronization stage. They properly account
for the energy-momentum conservation laws. One gets all possible exclusive
characteristics at a given energy but fails to learn their asymptotics. On the
contrary, analytical QCD pretends to start with asymptotical values and proceed
to lower energies accounting for conservation laws, higher order perturbative
and simplified non-perturbative effects.

The perturbative evolution is terminated at some low scale $Q_0\sim 1$GeV for
transverse momenta or virtualities of partons. Some observed variables, e.g.,
such as thrust, are insensitive to this "infrared" cut-off. For others, like
inclusive distributions, the local parton-hadron duality (LPHD) is assumed which
declares that the distributions at the parton level describe the hadron
observables up to some constant factor. This concept originates from the
preconfinement property of quarks and gluons to form colorless clusters.
It works surprisingly well when applied  for comparison with experiment.

Even more amazing look two other features of the perturbative QCD approach:
probabilistic description and applicability to comparatively soft processes.
At high energies, the quantum interference of different amplitudes for parton
production results in angular (or, more precisely, transverse momentum) 
ordering of successive emissions of gluons which favors the probabilistic
equations for these processes. The solutions of these equations obtained via the
modified perturbative expansion seem to be applicable sometimes even for rather
soft processes where the expansion parameter is not small enough and, moreover,
it is multiplied by some large factors increasing with energy. Thus, it can be
justified only because some subseries of the purely perturbative expansion 
ordered according to their high energy behavior are summed first and then the 
asymptotic series is cut off at the proper order.
In this framework, the perturbative QCD has demonstrated its very high 
predictive power.

If asked to choose 10 most spectacular analytical predictions
already confirmed by experiment I would mention:

1. the energy dependence of mean multiplicities,

2. oscillations of cumulant moments of multiplicity distributions as functions
of their rank,

3. difference between quark and gluon jets,

4. the hump-backed plateau of inclusive rapidity distribution and energy
dependence of its maxima,

5. difference between heavy- and light-quark jets,

6. color coherence in 3-jet events,

7. intermittency and fractality,

8. the energy behavior of higher moments of multiplicity distributions,

9. subjet multiplicities,

10. jet universality.

I have selected some of the most impressive results. More complete list with
a detailed survey can be found in the books \cite{1, dkmt} and in more recent
review papers \cite{dre1, koch, dgar, kowo}.

\section{QCD equations}

First, let me describe main theoretical tools used for prediction of all these
features. The most general approach starts from the equation for the generating
functional. The generating functional contains complete information about any
multiparticle process and is defined as
\begin{equation}
G(\{u\}, \kappa _0)=\sum _n\int d^3k_1...d^3k_n u(k_1)...u(k_n)P_n(k_1,...,k_n;
\kappa _0),
\end{equation}
where $P_n(k_1,...,k_n;\kappa _0)$ is the probability density for exclusive
production of particles with momenta $k_1,...,k_n$ at the initial virtuality
(energy) $\kappa _0$, and $u(k)$ is an auxiliary function. For $u(k)=$const,
one gets the generating function of the multiplicity distribution
$P_n(\kappa _0)$. The variations of $G(\{u\})$ over $u(k)$ (or differentials 
for constant $u$) provide any inclusive distributions and correlations of 
arbitrary order, i.e. complete information about the process. The general
structure of the equation for the generating functional describing the jet
evolution for a single species partons can be written symbolically as
\begin{equation}
G'\sim \int \alpha _SK[G\otimes G-G].
\end{equation}
It shows that the evolution of $G$ indicated by its variation (derivative)
$G'$ is determined by the cascade process of the production of two partons
by a highly virtual time-like parton (the term $G\otimes G$) and by the
escape of a single parton ($G$) from a given phase space region. The weights
are determined by the coupling strength $\alpha _S$ and the splitting function
$K$ defined by the interaction Lagrangian. The integral runs over all internal
variables, and the symbol $\otimes $ shows that the two partons share the
momentum of their parent. This is a non-linear integrodifferential probabilistic
equation with shifted arguments in the $G\otimes G$ term under the integral sign.

For quark and gluon jets, one writes down the system of two coupled equations.
Their solutions give all characteristics of quark and gluon jets and allow for
the comparison with experiment to be done. Let us write them down explicitly
for the generating functions now.
\begin{eqnarray}
&G_{G}^{\prime }&= \int_{0}^{1}dxK_{G}^{G}(x)\gamma _{0}^{2}[G_{G}(y+\ln x)G_{G}
(y+\ln (1-x)) - G_{G}(y)] \nonumber \\ 
&+&n_{f}\int _{0}^{1}dxK_{G}^{F}(x)\gamma _{0}^{2}
[G_{F}(y+\ln x)G_{F}(y+\ln (1-x)) - G_{G}(y)] ,   \label{50}
\end{eqnarray}
\begin{equation}
G_{F}^{\prime } = \int _{0}^{1}dxK_{F}^{G}(x)\gamma _{0}^{2}[G_{G}(y+\ln x)
G_{F}(y+\ln (1-x)) - G_{F}(y)] ,                                   \label{51}
\end{equation}
where $G^{\prime }(y)=dG/dy ,$
$y=\ln (p\Theta /Q_0 )=\ln (2Q/Q_{0}), p$ is an initial momentum, $\Theta $ 
is the angle of the divergence of the jet (jet opening angle), assumed here to be 
small, $Q$ is the jet virtuality,  $Q_{0}=$ const , 
$ n_f$ is the number of active flavours,
\begin{equation}
\gamma _{0}^{2} =\frac {2N_{c}\alpha _S}{\pi } ,                \label{52}
\end{equation}
the labels $G$ and $F$ correspond to gluons and quarks,
and the kernels of the equations are
\begin{equation}
K_{G}^{G}(x) = \frac {1}{x} - (1-x)[2-x(1-x)] ,    \label{53}
\end{equation}
\begin{equation}
K_{G}^{F}(x) = \frac {1}{4N_c}[x^{2}+(1-x)^{2}] ,  \label{54}
\end{equation}
\begin{equation}
K_{F}^{G}(x) = \frac {C_F}{N_c}\left[ \frac {1}{x}-1+\frac {x}{2}\right] ,   
\label{55}
\end{equation}
where $N_c$=3 is the number of colours, and $C_{F}=(N_{c}^{2}-1)/2N_{c}
=4/3$ in QCD. The asymmetric form (\ref{53}) of the three-gluon vertex can be 
used due to symmetry properties of the whole expression.
The variable $u$ has been omitted in the generating functions.

Let us note that these equations can be exactly solved \cite{21}
if the coupling strength is assumed fixed, i.e. independent of $y$.
For the running coupling strength, the Taylor series expansion can be
used \cite{13}
to get the perturbative expansion of physically measurable quantities.

A typical feature of any field theory with a dimensionless coupling constant
(quantum chromodynamics in particular) is the presence of the singular
terms at $x\rightarrow 0$ in the kernels of the equations. They imply the
uneven sharing of energy between newly created jets and play an important role
in jet evolution giving rise to its more intensive development compared
with the equal proportions (nonsingular) case.

Even though the system of equations (\ref{50}), (\ref{51}) is physically
appealing, it is not absolutely exact; i.e. it is not derived from first
principles of quantum chromodynamics. One immediately notices this since,
for example, there is no
four-gluon interaction term which is contained in the lagrangian of QCD.
Such a term does not contribute a singularity to the kernels and its
omission is justified in the lowest orders. Nevertheless, the modified series of
the perturbation theory (with three-parton vertices) is well
reproduced by such equations up to the terms including two-loop and
three-loop corrections. As shown in Ref. \cite{dkmt}, the neglected
terms would contribute at the level of the product of, at least, five generating
functions. Physical interpretation of the corresponding graphs would lead to
treatment of the 'colour polarizability' of jets. Sometimes, the effective 
infrared-safe coupling constant is used as a substitute for the 
phenomenological parameter $Q_0$. It must be universal 
for different processes and tend to a constant limit at low virtualities. 
However, we will use the more traditional approach. 

There are some problems with
the definition of the evolution parameter, with preasymptotic corrections etc.
 For example, the lower and upper limits of integration over $x$ in
 Eqns (\ref{50}), (\ref{51}) change in the preasymptotical region. It is 
imposed by the restriction on the transverse momentum which is given by
\begin{equation}
k_t = x(1-x)p\Theta ^{\prime } > Q_0/2.   \label{ktli}
\end{equation}
This condition originates from the requirement that the formation time of a
gluon  ($t_{form}\sim k/k_{t}^{2}$) should be less than its hadronisation time
($t_{had}\sim kR^{2}\sim k/Q_{0}^{2}$) for the perturbative QCD to be 
applicable. It leads to the requirement that the arguments of the generating
functions in Eqns (\ref{50}), (\ref{51}) should be positive. Therefore,
we must integrate in Eqns (\ref{50}), (\ref{51}) over 
$x$ from $\exp (-y)$ to $1-\exp (-y)$. However these limits tend to 0 and 1 at 
high energies. That is why it seems reasonable to learn more about the solutions
of the equations (\ref{50}), (\ref{51}) in the asymptotical region, and then
take the neglected terms into account as corrections to these solutions.

Moreover, it is of physics importance. With limits equal to $\exp (-y)$ and
$1-\exp (-y)$, the partonic cascade terminates at the perturbative level $Q_0/2$
as is seen from the arguments of the generating functions in the integrals.
With limits equal to 0 and 1, one extends the cascade into the non-perturbative
region with low virtualities $Q_1\approx xp\Theta /2$ and
$Q_2\approx (1-x)p\Theta /2$ less than $Q_0/2$. This region contributes terms of 
the order $\exp (-y)$, power-suppressed in energy. It is not clear whether the
equations and LPHD hypothesis are valid down to some $Q_0$ 
only or the non-perturbative region can be included as well.

Some approximations are used to solve these equations. The asymptotical results
are obtained in the so-called double-logarithmic (DLA) or leading order (LO)
approximation when the terms $(\alpha _S\ln ^2s)^n$ are summed. Here $s$ is 
the cms energy squared. The emitted gluons are assumed so soft that the
energy-momentum conservation is neglected. The corrections accounting for
conservation laws in the $G\otimes G$ term and in limits of the integration
as well as the higher order terms in the weight $\alpha _SK$ (in particular,
the non-singular terms of the kernels $K$) appear in the
next-to-leading (NLO or MLLA - modified leading logarithm approximation) and 
higher (2NLO,...) orders. Formally, these equations have been proven only for
the next-to-leading (NLO) order of the perturbative QCD. However, one can try
to consider them as kinetic equations in higher orders and/or generalize them
including the abovementioned effects in a more rigorous way than it is usually
implied.

\section{Comparison with experiment}

Let us turn directly to the comparison of results obtained with available
experimental data. The main bulk of the data is provided by
$e^{+}e^{-}$-processes at $Z^0$ energy.

\subsection{The energy dependence of mean multiplicity}

The equations for the average multiplicities in jets are obtained from
the system of equations (\ref{50}), (\ref{51}) by expanding the generating
functions in $u-1$ and keeping the terms
with $q$=0 and 1 with account of the definition
\begin{equation}
\frac{dG}{du}\vline _{u=1}=\sum nP_n=\langle n\rangle.
\end{equation}
They read
\begin{eqnarray}
\langle n_G(y)\rangle ^{'} =\int dx\gamma _{0}^{2}[K_{G}^{G}(x)
(\langle n_G(y+\ln x)\rangle +\langle n_G(y+\ln (1-x)\rangle -\langle n_G(y)
\rangle ) \nonumber  \\
+n_{f}K_{G}^{F}(x)(\langle n_F(y+\ln x)\rangle +\langle n_F(y+
\ln (1-x)\rangle -\langle n_G(y)\rangle )],  \label{ng}
\end{eqnarray}
\begin{equation}
\langle n_F(y)\rangle ^{'} =\int dx\gamma _{0}^{2}K_{F}^{G}(x)
(\langle n_G(y+\ln x)\rangle +\langle n_F(y+\ln (1-x)\rangle -\langle n_F(y)
\rangle ).   \label{nq}
\end{equation}
Herefrom one can learn about the energy evolution
of the ratio of multiplicities $r$ and of the QCD anomalous dimension $\gamma $
(the slope of the logarithm of average multiplicity in a gluon jet) defined as
\begin{equation}
r=\frac {\langle n_G\rangle }{\langle n_F\rangle }\; ,\;\;\;\;\; \;\;\;
\gamma =\frac {\langle n_G\rangle ^{'}}{\langle n_G\rangle }
=(\ln \langle n_G\rangle )^{'}\; .  \label{def}
\end{equation}
They have been represented
by the perturbative expansion at large $y$ as
\begin{equation}
\gamma = \gamma _{0}(1-a_{1}\gamma _{0}-a_{2}\gamma _{0}^{2}-a_3\gamma _0^3)+O(\gamma _{0}^{5})
 , \label{X}
\end{equation}
\begin{equation}
r = r_0 (1-r_{1}\gamma _{0}-r_{2}\gamma _{0}^{2}-r_3\gamma _0^3)+O(\gamma _{0}^{4})
.  \label{Y}
\end{equation}
Using the Taylor series expansion of $\langle n\rangle $ at large $y$ in
Eqns (\ref{ng}), (\ref{nq}) with (\ref{X}), (\ref{Y}) one gets the
coefficients $a_i,\, r_i$.

One of the most spectacular predictions of QCD states that in the leading order
approximation, where $\gamma =\gamma _0$, average multiplicities should
increase with energy \cite{mu1, dfkh, bcmm} like
$\exp [c\sqrt {\ln s}]$, i.e., in between the power-like and logarithmic
dependences predicted by hydrodynamical and multiperipheral models.
Next-to-leading order results account for the term with $a_1$ in Eqn. (\ref{X})
\cite{web1, dktr, cdfw} and contribute the logarithmically decreasing
factor to this behavior whereas the higher order terms do not practically
change this dependence \cite{dg, cdnt8}. The fitted parameters in the final
expression are an
overall constant normalization factor which is defined by confinement and a
scale parameter $Q_0$. The $e^{+}e^{-}$-data are well fitted by such an
expression as seen in Fig. 1. Let us note here that the expansion parameter
$\gamma $ is rather large at present energies being about 0.4 - 0.5.

\subsection{Oscillations of cumulant moments} 

The shape of the multiplicity distribution can be described by its higher
moments related to the width, the skewness, the kurtosis etc. The $q$-th
derivative of the generating function corresponds to the factorial moment
$F_q$, and the derivative of its logarithm defines the so-called cumulant
moment $K_q$. The latter ones describe the genuine (irreducible) correlations
in the system (it reminds the connected Feynman graphs).
\begin{equation}
F_{q} = \frac {\sum_{n} P_{n}n(n-1)...(n-q+1)}{(\sum_{n} P_{n}n)^{q}} =
\frac {1}{\langle n \rangle ^{q}}\cdot \frac {d^{q}G(z)}{du^{q}}\vline _{u=1}, 
\label{4}
\end{equation}
\begin{equation}
K_{q} = \frac {1}{\langle n \rangle ^{q}}\cdot \frac {d^{q}\ln G(z)}{du^{q}}
\vline _{u=1}, \label{5}
\end{equation}
where
\begin{equation}
\langle n \rangle = \sum_{n=0}^{\infty }P_{n}n              \label{6}
\end{equation}
is the average multiplicity.
These moments are not independent. They are connected by definite relations
that can easily be derived
from their definitions in terms of the generating function:
\begin{equation}
F_{q} = \sum _{m=0}^{q-1} C_{q-1}^{m} K_{q-m} F_{m} ,              \label{11}
\end{equation}
which are nothing other than the relations between the derivatives of a function
and of its logarithm at the point where the function itself equals 1. Here
\begin{equation}
C_{q-1}^{m} = \frac {(q-1)!}{m!(q-m-1)!} = \frac {\Gamma (q)}{\Gamma (m+1)
\Gamma (q-m)} = \frac {1}{mB(q,m)}     \label{12}
\end{equation}
  are the binomial coefficients, and $\Gamma $ and $B$ denote the gamma-  and
 beta-functions, correspondingly. Thus there are only numerical
  coefficients in the recurrence relations (\ref{11}) and the iterative
solution (well-suited for computer calculation) reproduces all cumulants if
the factorial moments have been given, and vice versa. In that sense, cumulants
and factorial moments are equally suitable. The relations for the low ranks are
\begin{eqnarray}
F_{1}&=&K_{1}=1, \nonumber  \\
F_{2}&=&K_{2}+1, \nonumber  \\
F_{3}&=&K_{3}+3K_{2}+1.   \label{12a}
\end{eqnarray}

Solving the Eqns. (\ref{50}), (\ref{51}), one gets quite naturally the
predictions \cite{13, 21, 41} for the behavior of the ratio $H_q=K_q/F_q$.
At asymptotically high energies, this ratio is predicted to behave as $q^{-2}$.
However, the asymptotics is very far from our realm.
At present energies, according to QCD, this ratio should reveal the minimum
at $q\approx 5$ and subsequent oscillations. This astonishing qualitative
prediction has been confirmed in experiment (for the first time in Ref.
\cite{dabg})
as in $e^{+}e^{-}$ (see Fig. 2 from \cite{sld})
as in hadronic processes. The predicted negative minimum of $H_q$ is clearly
observed. It can correspond to the replacement of attractive
forces (clustering) by repulsion (between clusters) in systems with different
number of particles. The minimum position slowly changes with energy and with
size of the phase space window. Also, for instanton induced processes it has
been found at $q\approx 2$.

The quantitative analytical estimates are not enough
accurate because, first of all, the expansion parameter becomes equal to the
product $q\gamma $ which is close to one or even exceeds it for all $q>1$.
Therefore the perturbative approach is, strictly speaking, inapplicable to
this problem. However, some tricks can be used to improve it. At the same time,
the numerical computer solution \cite{lo2} reproduces oscillations quite well.
These new laws differ from all previously attempted distributions of the
probability theory. 

\subsection{Difference between quark and gluon jets}

The system of two equations for quark and gluon jets predicts that
asymptotically the energy dependence of mean multiplicities in them should
be identical. However, normalization differs, and gluon jets are more "active"
so that the ratio
$r=\langle n_G\rangle /\langle n_F\rangle $ of average multiplicities in
gluon and quark jets should tend at high energies \cite{brgu}
to the ratio of Casimir operators $C_A/C_F=9/4$. Once again,
this prediction shows how far are we now from the true asymptotics because
in experiment this ratio is about 1.5 at $Z^0$ energy and even smaller at
lower energies. The higher order terms \cite{43, web1, cdnt8}
 (calculated now up to 3NLO) improve the agreement and approach the experimental value with an accuracy about 15$\%$
(see Fig. 3). The higher order terms change slightly also the energy behavior
of quark jets compared to gluon jets as observed in experiment. However, the
simultaneous fit of quark and gluon jets with the same set of fitted
parameters is still not very accurate as is seen from the shaded area in
Fig. 1. This failure is again due to insufficiently precise description
of the ratio $r$.

Even better agreement has been achieved when the equations are
modified to account for phase space limitations imposed by energy-momentum
conservation \cite{eden}. However, some problems arise for higher moments of the
multiplicity distribution in such an analytical approach \cite{dede}. The exact computer
solution of the equations \cite{lo1} has led to better agreement on the
ratio $r$ at $Z^0$
but about 20$\%$ discrepancy is still left at lower energies of $\Upsilon $. 

The widths of the multiplicity distributions differ in quark and gluon jets,
the former being somewhat wider. Qualitatively, QCD describes this tendency
but quantitative estimates are rather uncertain yet as is discussed in the
subsection 3.8.

\subsection{The hump-backed plateau}

Dealing with inclusive distributions, one should solve the equations for the
generating functional. It has been done up to NLO approximation.
As predicted by QCD, the momentum (rapidity $y$) spectra of particles inside
jets should have the shape of the hump-backed plateau \cite{dfkh, bcmm, adkt}.
This striking prediction of the perturbative QCD differs from the previously
popular flat plateau advocated by Feynman.
It has been found in experiment (Fig. 4).
The depletion between the two humps is due to angular ordering and color
coherence in QCD. The humps are of the approximately Gaussian shape near
their maxima if the variable
\begin{equation}
\xi=\ln \frac {1}{x}; \;\;\;\;\; x=\frac {p}{E_j}
\end{equation}
is used. Here $p$ is the particle momentum, $E_j$ is the jet energy. This
prediction was first obtained in the LO QCD, and more accurate expressions were
derived in NLO \cite{fweb}. Moments of the distributions up to the fourth rank
have been calculated. The drop of the spectrum towards small momenta becomes more
noticeable in this variable. The comparison with experimental data at different
energies has revealed good agreement both on the shape of the spectrum (see
Fig. 5 for $e^{+}e^{-}$ from \cite{delp}) and on the energy dependence of its peak position 
(see Fig. 6 for $e^{+}e^{-}, ep, p\overline p$ from \cite{cdf}) and width.

\subsection{Difference between heavy- and light-quark jets}

Another spectacular prediction of QCD is the difference between the spectra
and multiplicities in jets initiated by heavy and light quarks. Qualitatively,
it corresponds to the difference in bremsstrahlung by muons and electrons 
where the photon emission at small angles is strongly suppressed for muons
because of the large mass in the muon propagator. Therefore, the intensity of
the radiation is lower in the ratio of masses squared. The coherence of soft
gluons also plays an important role in QCD. For heavy quarks the accompanying
radiation of gluons should be stronger depleted in the forward direction
(dead-cone or ring-like emission). It was predicted \cite{83, sdkk} that it should
result in the energy-independent difference of companion mean multiplicities
for heavy- and light-quark jets of equal energy. The naive model of energy
rescaling \cite{85, 86, 87} predicts the decreasing difference. The experimental
data (see Fig. 7 from \cite{del1}) support this QCD conclusion.

\subsection{Color coherence in 3-jet events}  

When three or more partons are involved in a hard interaction, one should take 
into account color-coherence effects. Several of them have been observed.
In particular, the multiplicity can not
be represented simply as a sum of flows from independent partons. QCD predicts
that the particle flows should be enlarged in the directions of emission of
partons and suppressed in between them. Especially interesting is the prediction 
that this suppression is stronger between $q\overline q$-pair than between
$gq$ and $g\overline q$ in $e^{+}e^{-}\rightarrow q\overline {q}g$ event if all
angles between partons are large (the "string" \cite{151} or "drag" \cite{64}
effect). All these predictions have been confirmed by experiment (see Fig. 8
from \cite{del2}). In $q\overline qg$ events the particle population values
in the $qg$ valleys are found larger than in the $q\overline q$ valley by a
factor 2.23$\pm $0.37 compared to the theoretical prediction of 2.4. Moreover,
QCD predicts that this shape is energy-independent up to an overall
normalization factor.

Let us note that for the process $e^{+}e^{-}\rightarrow q\overline q\gamma $
the emission of additional photons would be suppressed both in the direction of
a primary photon and in the opposite one. In the case of an emitted gluon, we
observe the string (drag) effect of enlarged multiplicity in its direction and
stronger suppression in the opposite one. This suppression is described by the
ratio of the corresponding multiplicities in the $q\overline q$ region
\begin{equation}
R_{\gamma }=\frac {N_{q\overline q}(q\overline qg)}
{N_{q\overline q}(q\overline q\gamma )}
\end{equation}
which is found to be equal 0.58$\pm $0.06 in experiment whereas the theoretical
prediction is 0.61.

The color coherence reveals itself as inside jets as in inter-jet regions.
It should suppress both the total multiplicity of
$q\overline {q}g$ events and the particle yield in the transverse to the
$q\overline {q}g$ plane for decreasing opening angle between the low-energy
jets. When hard gluon becomes softer, color coherence determines, e.g., the
azimuthal correlations of two gluons in $q\overline qgg$ system. In particular,
back-to-back configuration ($\varphi \sim 180^0$) is suppressed by a factor
$\sim 0.785$ in experiment, 0.8 in HERWIG Monte Carlo and 0.93 in analytical
pQCD. In conclusion, color coherence determines topological dependence of jet
properties.

Results for 
jets in $ep$ and $p\overline p$ processes also favor theoretical expectations.
Some proposals have been promoted for a special two-scale analysis of 3-jet
events when the restriction on the transverse momentum of a gluon jet is 
imposed \cite{egus, egkh}. They are under experimental study now.

\subsection{Intermittency and fractality}

The self-similar parton cascade leads to special multiparton correlations.
Its structure with "jets inside jets inside jets..." provoked the analogy with
turbulence and the ideas of intermittency \cite{66}. Such a structure should
result in the fractal distribution in the available phase space \cite{drje}.
The fractal behavior would display the linear dependence of logarithms of
factorial moments on the logarithmic size of phase space windows. The moments
are larger in smaller windows, i.e. the fluctuations increase in smaller bins
in a power-like manner (see the review paper \cite{14}).

In QCD, the power dependence appears for a fixed coupling regime \cite{21}.
The running property of the coupling strength in QCD flattens \cite{38, 68, 70}
this dependence at
smaller bins, i.e. the multifractal behavior takes over there. The slopes for
different ranks $q$ are related to the Renyi dimensions. Both the linear
increase at comparatively large but decreasing bins and its flattening for
small bins have been observed in experiment (see Fig. 9 from \cite{opal}). 
However, only
qualitative agreement with analytical predictions can be claimed here. The
higher order calculations are rather complicated and mostly the results
of LO with some NLO corrections are yet available. In experiment, different
cuts have been used which hamper the direct comparison. However, Monte Carlo
models where these cuts can be done agree with experiment better. The role of
partonic and hadronization stages in this regime is still debatable.

\subsection{The energy behavior of higher moments of multiplicity distributions}

The factorial moments increase both with their rank and with energy increasing.
From the mentioned above behavior of $H_q$-moments one easily guesses that the
the cumulant moments behave in a similar but somewhat different manner.
The experimental results for 41.8 GeV gluon jets $F_2^G=1.023$ and
for 45.6 GeV $uds$ quark jets, $F_2^F=1.082$ are
much smaller than the asymptotical predictions, viz. 1.33 and 1.75,
respectively. The NLO terms improve the description of the data compared to
leading order results. If one accepts the effective value of $\alpha _S$ 
averaged over all the energies of the partons during the jet evolution to be
$\alpha _S\approx 0.2$, one obtains the NLO values $F_2^G\approx 1.039$ and
$F_2^F\approx 1.068$ at these energies which are quite close to experimental
results. In this sense the NLO prediction can be said to describe the widths
of the gluon and quark jet multiplicity distributions at $Z^0$ energy to within
10$\%$ accuracy.

Unfortunately, the 2NLO and 3NLO terms worsen the agreement with data compared
to NLO (but not compared to LO) results \cite{dlne}.
The same is true for higher order
moments. It raises the general problem of the convergence of the perturbative
expansion in view of the large expansion parameter $q\gamma $ mentioned above.
The attempts to account for conservation laws more accurately by
the modified evolution equations for high moments \cite{dede} have not led to
the success yet. It is remarkable that the computer solution of the QCD
equations \cite{lo2} provides a
near-perfect description of the higher moments as well. This suggests that
the failure of the analytic approach at higher orders is mainly a technical 
issue related to a treatment of soft gluons.

\subsection{Subjet multiplicities}

A single quark-antiquark pair is initially created in $e^{+}e^{-}$-annihilation.
With very low angular resolution one observes two jets. A three-jet structure
can be observed when a gluon with large transverse momentum is emitted by the
quark or antiquark. However such a process is suppressed by an additional
factor $\alpha _S$, which is small for large transferred momenta. It can be
calculated perturbatively. At relatively low transferred momenta, the jet
evolves to angular ordered subjets ("jets inside jets...").
Different algorithms have been proposed to resolve subjets. By increasing the
resolution, more and more subjets are observed. For very high resolution, the
final hadrons are resolved. The resolution criteria are chosen to provide
infrared  safe results.

In particular, one can predict the asymptotical ratio of subjet multiplicities
in 3- and 2-jet events if one neglects soft gluon coherence:
\begin{equation}
\frac {n_3^{sj}}{n_2^{sj}}=\frac {2C_F+C_A}{2C_F}=\frac {17}{8}.
\end{equation}
Actually, the coherence reduces this value to be below 1.5 in experiment for
all acceptable resolution parameters. Theoretical predictions \cite{cdfw}
agree only qualitatively with experimental findings \cite{l3, opa1}.

Subjet multiplicities have also been studied in separated quark and gluon jets.
The analytical results \cite{seym} are seen (Fig. 10 from \cite{alep}) to
represent the data fairly well for
large values of the subjet resolution scale $y_0$.

\subsection{Jet universality}

According to QCD, jets produced in processes initiated by different colliding
particles should be universal and depend only on their own parent (gluon, light
or heavy quark). This prediction has been confirmed by many experiments. In this
talk, e.g., it is mentioned in subsections 3.2, 3.6, 3.7 and
demonstrated in Fig. 6.

\section{Conclusions and outlook}
 
A list of successful analytical QCD predictions can be made longer. Quantum
chromodynamics has already predicted spectacular qualitative features of soft
processes. Quantitatively, analytical results show that higher order (NLO) terms
always tend to improve the agreement with experiment compared to asymptotical
(LO) predictions. The accuracy achieved is often better than 20$\%$ or even
10$\%$ that is surprising by itself considering rather large expansion parameter
of the perturbative approach. Moreover, some characteristics are very sensitive
to ever higher order terms and should be carefully studied. The astonishing
success of the computer solutions and Monte Carlo schemes demonstrates the
importance of the treatment of the boundary between the
perturbative and non-perturbative regions which is approximately taken into
account in the analytical approach via the cut-off parameter $Q_0$ and
the limits of integration over the parton splitting variables.\\

A new era of multiparticle production studies opens with the advent of RHIC,
LHC, TESLA. We are coming closer to the asymptotic region even though the 
predicted dependences are very slow. Nevertheless, some results differ for
various analytical approaches and Monte Carlo schemes at these energies.
It will allow to distinguish between them.

The mean multiplicities will increase drastically. Now, in Au-Au collisions
at 130 GeV per nucleon at RHIC the mean charged multiplicity exceeds 4000.
It implies that the event-by-event analysis of various patterns formed by
particles in the available phase space becomes meaningful. The results
can be compared to exclusive probabilistic Monte Carlo schemes. The qualitative
QCD predictions indicate the tendencies towards the asymptotical region.
It would allow to analyze small color-suppressed effects, properties of minijets 
or clusters (with attraction-repulsion transition), other collective effects
like elliptic flow (and even higher Fourier expansion terms), ring-like events
(the probable signature of "Cherenkov gluons") etc. The event-by-event analysis
of experimental exclusive data can be quantified locally if one uses wavelets 
(sometimes called "mathematical microscope") for pattern recognition in
individual events \cite{dine}. We hope to confront QCD predictions with new
findings as
well as to separate new physics signals from conventional QCD background.\\

\end{document}